\documentclass[useAMS,usenatbib,usegraphicx]{mn2e}
\pubyear{2010}

\newcommand{\be}{\begin{equation}}
\newcommand{\ee}{\end{equation}}

\begin{document}

\title[Re-analysis of WMAP 7 year cosmological parameters]
{Cosmological Parameters from a re-analysis \, \\ of the WMAP 7 year low resolution maps}
\author[F. Finelli, A. De Rosa, A. Gruppuso, D. Paoletti]
{F.~Finelli $^{1,2}$\thanks{E-mail: finelli@iasfbo.inaf.it }, A. De Rosa $^{1}$\thanks{E-mail: derosa@iasfbo.inaf.it}, 
A.~Gruppuso $^{1,2}$\thanks{E-mail: gruppuso@iasfbo.inaf.it }, D. Paoletti $^{1,2}$\thanks{E-mail: paoletti@iasfbo.inaf.it}
\\
$^1$ INAF-IASF Bologna, Istituto di Astrofisica Spaziale e Fisica Cosmica 
di Bologna \\
Istituto Nazionale di Astrofisica, via Gobetti 101, I-40129 Bologna, Italy \\
$^2$ INFN, Sezione di Bologna,
Via Irnerio 46, I-40126 Bologna, Italy}

\label{firstpage}

\maketitle

\begin{abstract}
Cosmological parameters from WMAP 7 year data are re-analyzed by substituting 
a pixel-based likelihood estimator to the one delivered publicly by the WMAP team. 
Our pixel based estimator handles exactly 
intensity and polarization in a joint manner, allowing to use low-resolution maps and noise covariance matrices 
in $T,Q,U$ at the same resolution, which in this work is 3.6$^\circ$. 
%$N_{\rm side}=16$.
We describe the features and the performances of the code implementing our pixel-based likelihood estimator. 
We perform a battery of tests on the application of our pixel based likelihood 
routine to WMAP publicly available low resolution foreground cleaned products, in combination with the 
WMAP high-$\ell$ likelihood, reporting the differences on cosmological parameters evaluated 
by the full WMAP likelihood public package. The differences are not only due to the treatment of polarization, but also 
to the marginalization over monopole and dipole uncertainties present in the WMAP pixel likelihood code for temperature. 
The credible central value for the cosmological parameters change below the 1 
$\sigma$ level with respect to the evaluation by the full WMAP 7 year likelihood code, with the largest difference in a shift 
to smaller values of the scalar spectral index $n_S$. 
\end{abstract}

%\label{firstpage}


\section{Introduction}

The anisotropy pattern of the cosmic microwave background (CMB) is a treasure for 
understanding the costituents of our Universe and how it evolved from the Big Bang.
Under the assumption of isotropy and Gaussianity of CMB fluctuations, 
the power spectra of intensity and polarization anisotropies include all the compressed information on our Universe 
through the determination of the cosmological parameters. 
There has been a tremendous improvement in the estimate of cosmological parameters driven by the increasingly better quality of CMB data, 
mainly due to the full sky observations in temperature and polarization by the Wilkinson Microwave Anisotropy Probe (WMAP),
(see \cite{Larson:2010gs,Komatsu:2010fb} and references therein) and to the small angular scales measurements by 
QUaD in polarization 
\citep{QUAD}, by the South Pole Telescope \citep{SPT,SPT2011,SPT2011_2} and 
the Atacama Cosmology Telescope \citep{ACT,ACT2010} in temperature.
{\sc Planck} will lead to a drastic improvement of CMB full sky maps in temperature and polarization, 
leading to an eagerly expected improvement in cosmological parameters with uncertainties at the percent level \citep{bluebook}.

A joint likelihood analysis in temperature and polarization is 
one of the accepted methods in securing the scientific expectations of observational achievements in terms of cosmological parameters.
Although the likelihood could be written exactly in the map domain under the Gaussian hypothesis, its computation is almost prohibitive 
already at the resolution of 2 degrees, whereas cosmological information is encoded in the temperature and polarization 
power spectra up to the angular scales of the order of few arcminutes, where the Silk damping suppress the CMB primary anisotropy spectrum. 
It is now commonly accepted to use an hybrid approach 
which combines a pixel approach at low resolution with an approximated likelihood based on power spectrum estimates at high multipoles (see 
\cite{BJK,verdeetal,HL} for some of these approximations).  

Since the three year release of the full polarization information, 
the WMAP team adopted such a hybrid scheme approach,  
%for a joint likelihood combining a pixel approach to the exact 
%likelihood on low resolution maps with an approximated likelihood based on power spectrum estimates at high multipoles. 
%Such hybrid approach 
which has been suggested independently in \cite{Efstathiou04,Slosar04,ODwyer04,Efstathiou06}.
At a first appearance of the three year data, the WMAP team adopted a pixel approach on 
HEALPIX \citep{gorski} resolution $N_{\rm side}=8$ \footnote{The number of pixels in a map is given by 
$N_{\rm pix} = 12 N_{\rm side}^2$, i.e. 768 for $N_{\rm side}=8$ and 3072 for $N_{\rm side}=16$.} 
temperature and polarization maps, and considered  
the high-$\ell$ approximated likelihood to start at $\ell =13$ in temperature and $\ell=24$ in polarization and 
temperature-polarization cross-correlation 
for the determination of cosmological parameters in \cite{Spergel:2006hy}. 
The WMAP team treats separately temperature and polarization as explained in \cite{Page:2006hz} 
and \cite{Hinshaw:2006ia}, 
by using the approximation that the noise in temperature is negligible. As a consequence, 
the WMAP likelihood code includes either $(Q,U)$ and the 
temperature-polarization cross-correlation in the same sub-matrix.
It was then shown by \cite{Eriksenetal2007} that by increasing the resolution of the temperature map to HEALPIX 
$N_{\rm side}=16$ and therefore the multipole of transition to high-$\ell$ approximated likelihood 
in temperature from $\ell=12$ to $\ell=30$, the mean value for the scalar spectral index $n_s$ 
shifted to higher values by a 0.4 $\sigma$. The asymmetric handling of the low-resolution 
temperature map at $N_{\rm side}=16$ and polarization at $N_{\rm side}=8$, 
%allowed by separate treatment of temperature from polarization and temperature-polarization cross-correlation, 
became the final treatment of the three year data release.
This low-$\ell$ likelihood aspect in the WMAP hybrid approach has not changed since the final release 
of the WMAP 3 year data to the current WMAP 7 year one.

In this paper we wish to perform an alternative determination of the cosmological parameters from WMAP 7 public data, substituting the 
WMAP low-$\ell$ likelihood approach with a pixel based likelihood code which treats $T,Q,U$ at the same HEALPIX resolution $N_{\rm side}=16$ 
connected to the standard WMAP high-$\ell$ package. In this analysis we therefore increase the resolution of polarization products digested 
by the pixel base likelihood from 
$N_{\rm side}=8$ to $N_{\rm side}=16$, in analogy with what done by \cite{Eriksenetal2007} for temperature only.
The WMAP 7 year foreground cleaned $(Q,U)$ maps, covariance matrices and masks 
at the resolution $N_{\rm side}=16$ are also publicly available at http://lambda.gsfc.nasa.gov: therefore, all data used in this paper 
are made available by the WMAP team. 

The paper is organized as follows. In Section II we briefly describe the WMAP hybrid approach to the likelihood, with particular care to 
the low multipole part. In Section III we describe our pixel approach, implemented in the BoPix code. We then present in Section IV 
the cosmological parameters obtained by using our alternative pixel approach in place of the WMAP one for a $\Lambda$CDM scenario. 
In Section V we extend our investigations to other cosmological models. In Section VI we draw our conclusions.

\section{A Brief Description of the WMAP Hybrid Likelihood Analysis}
\label{description}

In the map domain, the likelihood as function of the cosmological parameters $\{ \theta \}$
\begin{equation}\label{eq:likelihood}
{\cal L} (\mathbf{d}|{\theta}) = \frac{1}{\left|2\pi \mathbf{C}\right|^{1/2}}\exp\left[-\frac12 
\mathbf{d}^t \mathbf{C}^{-1} \mathbf{d}\right]
\end{equation}
where the data, $\mathbf{d}=\mathbf{s}+\mathbf{n}$, is a CMB fully polarized map, 
considered as a vector combining $T$, $Q$ and $U$ foreground reduced maps, the sum of 
signal $\mathbf{s}$ and noise $\mathbf{n}$; the quantity $\mathbf{C}=\mathbf{S}+\mathbf{N}$ 
is the total covariance matrix, the sum of the CMB signal covariance matrix $\mathbf{S} (\theta)$, 
and the noise matrix $\mathbf{N}$. 
The signal covariance matrix is constructed by the power spectra $C^{XY}_\ell$, where $X,Y$ are 
any of $T,E,B$ \citep{Zaldarriaga:1996xe} 
as given in \cite{tegmark_pol}: if not otherwise stated, the sum over multipoles starts from $\ell=2$. 
%(as mentioned above, in the absence of partiy violation, the CMB only has $XY=TT, TE, EE, BB$).

The WMAP low-$\ell$ likelihood is described in the Appendix of \cite{Page:2006hz} and we report here the essentials.
The WMAP approach is based on the assumption to ignore the noise in temperature, which leads to a simplification of 
the likelihood, useful from the numerical computation perspective. 
By assuming that the noise in temperature is negligible at low multipoles, 
the WMAP approach consists in rewriting Eq. (\ref{eq:likelihood}) as:
\begin{eqnarray}
{\cal L} (\mathbf{d}|{\theta}) &\simeq& 
\frac{\exp\left(-\frac12 \mathbf{s}_T^t  S_{T}^{-1} \mathbf{s}_T \right)}{\sqrt{2 \pi} |S_{T}|^{1/2}} \times \nonumber \\
& & \frac{\exp\left[-\frac12  \tilde{\mathbf{d}}_P^t (\tilde{S}_P+\tilde{N}_P)^{-1}
\tilde{\mathbf{d}}_P \right]}{|\tilde{S}_P+N_P|^{1/2}} \,
\label{eq:dnslike}
\end{eqnarray}
where $S_T$ is the temperature signal sub-matrix, 
the new polarization data vector is $\tilde{\mathbf{d}}_P = \tilde{\mathbf{s}}_P + \tilde{\mathbf{n}}_P$, with 
$\tilde{\mathbf{s}}_P =(\tilde{Q},~\tilde{U})$ given by
\begin{eqnarray}
 \tilde{Q} &\equiv& Q - \frac12\sum_{l=2}^{\ell_P}
  \frac{C_l^{TE}}{C_l^{TT}} \sum_{m=-l}^l a_{\ell m}^{TT}({}_{+2}Y_{lm}+{}_{-2}Y^*_{lm}),
\label{eq:qtilde}
\\
 \tilde{U} &\equiv& U - \frac{i}2\sum_{\ell=2}^{\ell_P}
  \frac{C_\ell^{TE}}{C_\ell^{TT}} \sum_{m=-\ell}^\ell a_{\ell m}^{TT}
({}_{+2}Y_{\ell m}-{}_{-2}Y^*_{\ell m}),
\label{eq:utilde}
\end{eqnarray}
with $\tilde{S}_P$ ($\tilde{N}_P$) is the signal (noise) covariance matrix for the new polarization vector \citep{Page:2006hz}.
The noise covariance matrix for $(\tilde{Q},\tilde{U})$ equals the original one for $(Q,U)$ when the noise in temperature is zero 
\citep{Page:2006hz}. As temperature $a_{\ell m}^{TT}$, the full-sky internal linear
combination (ILC) map is used \citep{Hinshaw:2006ia}.
%To estimate $T_{lm}$, we used the full-sky internal linear 
%combination (ILC) temperature map \citep{hinshaw/etal:prep}. 

%It is important to stress 
According to \cite{Page:2006hz},  Eq.~(\ref{eq:likelihood}) and Eq.~(\ref{eq:dnslike})
are mathematically equivalent when the temperature noise is ignored.
With this assumption, the new form, Eq.~(\ref{eq:dnslike}), allows the WMAP approach to factorize
the likelihood of temperature and polarization, with
the information in their cross-correlation, 
$C_\ell^{TE}$, retained in the polarization sub-matrix.
As already mentioned in the introduction, temperature is considered at the HEALPIX resolution $N_{\rm side}=16$ and smoothed with a 
Gaussian beam of 9.1285$^\circ$, 
whereas polarization is considered at $N_{\rm side}=8$ and not smoothed.
The range of multipoles used in the polarization sub-matrix is up to the Nyquist 
limit at $N_{\rm side}=8$, i.e. $\ell_P=23$.
Two computation options are available for the temperature likelihood, 
Gibbs sampling \citep{jewell04,wandelt04,eriksen04} with a range of multipole considered up to $\ell_T=32$ 
and direct pixel evaluation, with $\ell_T=30$ \footnote{The temperature signal covariance matrix is constructed with 
multipoles up to $\ell=64$, but from $\ell=31$ to $64$ the $C_\ell^{TT}$ are not varied, but 
fixed to those of a fiducial cosmology.}.
All the computations by the WMAP low-$\ell$ likelihood 
reported here are performed with the option {\em ifore}=2 for temperature (we have checked that 
differences are minimal with respect to the alternative options 
{\em ifore}=0 and 1) and without considering marginalization over foreground uncertainties in polarization.
%\footnote{It is recommended 
%to use two different transition multipoles at which the high-$\ell$ temperature likelihood starts for these low-$\ell$ 
%computational options: $\ell=32$ for Gibbs sampling and $\ell=30$ for pixel evaluation.}.

The high-$\ell$ likelihood, described in \cite{Larson:2010gs} and in \cite{verdeetal}, 
has been updated to beam/point sources uncertainties through the various subsequent WMAP releases 
\citep{Hinshaw:2006ia,nolta}.
The high-$\ell$ TT likelihood takes into account multipoles from 
$\ell=31$ ($\ell=33$) when connected with the pixel (Gibbs) 
likelihood evaluation of the low resolution temperature data up to $\ell=1200$;   
the high-$\ell$ TE (and TB when used) likelihood takes into account multipoles 
from $\ell=24$ \citep{Page:2006hz} to $\ell=800$. 
High-$\ell$ EE and BB data have not used so far in the various relases of the WMAP likelihood code.

\section{BoPix}
\label{bopix}

BoPix computes the likelihood function in Eq. (\ref{eq:likelihood}) for the parameter
space $\{ \theta \}$ which the $C_\ell^{XY} (\{ \theta \})$ depend on, {\em without any 
approximation and with the same resolution in temperature and polarization}.
BoPix is a multithreaded OpenMP Fortran90 library
which can be connected to a sampler - to CosmoMC \cite{cosmomc} in this work.

The computation of the likelihood given in Eq. (1) requires an environment initialization, 
in which BoPix calculates the geometrical functions dependent on the cosine of the angle between two pixels
and reads the noise covariance matrix (C-binary format). 

BoPix then starts to compute the signal covariance matrix ${\bf S}$ for a given $C_\ell^{XY} (\{ \theta \})$ with a OpenMP routine 
with a high intrinsic 
level of parallel architecture, to which the noise covariance matrix ${\bf N}$ is summed. The full covariance matrix is then Cholesky 
decomposed. The computation of the determinant is obtained from 
the properties of the Cholesky decomposed matrix ${\bf L}$: ${\rm det} {\bf C} = ({\rm det} {\bf L})^2$. 
The term ${\bf C}^{-1} {\bf d}$ is computed as the solution for the variable ${\bf x}$ 
(vector with dimension $3 N_{\rm pix}$) of the equation ${\bf C x} = {\bf d}$.
 
The matrix manipulations are implemented on LAPACK and BLAS mathematical libraries (as nag, essl, acml and mkl).
There is an effort to improve the BoPix capabilities and performances (in
terms of run time and memory) to make the direct
likelihood evaluation at low resolution for cosmological parameters
extraction as fast as possible, in particular by reducing the time 
spent for the Cholesky decomposition, and optimizing the combined scalability in memory and CPU time of this code; 
indeed, the resources required by BoPix are larger than those 
for the WMAP low-$\ell$ likelihood code since the polarization sector is treated at higher resolution.
%Note that the algorithm summarized above does not include the computation of ${\bf C}^{-1}$. 
%The computation time required by BoPix is larger 
%than the corresponding one required by the WMAP likelihood low-$\ell$ code since the resolution in polarization is higher.
At present, BoPix can handle maps and full noise
covariances up to HEALPIX $N_{\rm side}=32$ resolution.
On IBM Power6 (4.2GHz) architecture, available at CINECA (http://www.cineca.it), with 64 threads on 64 logical CPUs (32 cores) BoPix
can calculate the likelihood in about 0.3 seconds at $N_{\rm side}=16$, and in about 15 seconds at $N_{\rm side}=32$.
At $N_{\rm side}=16$ on the same IBM Power6, a good trade off between computation time and memory required is obtained for 2 sec with 8 cores. 
More details about performances and comparison among different platforms will be provided in \cite{DeRosa2013}.

\section{Data set for BoPix}
\label{dataset}

%In this Section we describe the data set that we have considered. 
We use the temperature ILC map smoothed at $9.1285$ degrees and reconstructed at HealPix 
%\footnote{http://healpix.jpl.nasa.gov/}
\citep{gorski} resolution $N_{\rm side}=16$, 
the foreground cleaned (unsmoothed) low resolution maps and the noise covariance matrix in $(Q,U)$ publicly available at the LAMBDA website
http://lambda.gsfc.nasa.gov/ for the frequency channels Ka ($23$ GHz), Q ($41$GHz) and V ($61$ GHz) 
as considered by \cite{Larson:2010gs} for the low 
$\ell$ analysis. These frequency channels have been co-added by inverse noise covariance weigthing 
accordingly to the WMAP team \citep{Jarosik}
\be
{\bf d}_{\rm pol} = {\bf c}_{\rm pol} ({\bf c}_{Ka}^{-1} {\bf d}_{Ka} + {\bf c}_Q^{-1} {\bf d}_Q+ {\bf c}_V^{-1} {\bf d}_V)
\, ,
\ee
where ${\bf d}_{i}$, ${\bf c}_i$ are the foreground reduced polarization maps and 
covariances, respectively (for i=Ka, Q and V). The total foreground reduced inverse noise covariance matrix is therefore:
\be
{\bf c}_{\rm pol}^{-1} = {\bf c}_{Ka}^{-1} + {\bf c}_{Q}^{-1} +{\bf c}_{V}^{-1} 
\, .
\ee
This polarization data set has been extended to temperature considering the ILC map with an extra noise term, 
as suggested in \cite{dunkley_wmap5}. 
We have therefore added to the temperature map a random noise realization with variance of $\sigma_{TT}^2 = 1 \mu K^2$ and 
consistently, the noise covariance matrix for TT is taken to be diagonal with variance equal to $1 \mu K^2$.
The total noise covariance ${\bf N}$ for WMAP 7 yr data is therefore:
\[ 
{\bf N} =
\left( \begin{array}{cc}
\sigma_{TT}^2 {\bf I} & 0 \\
0 & {\bf c}_{\rm pol} \\
\end{array} \right)
\]
Let us note that this prescription of the noise in the temperature ILC map added to mitigate the uncertainties 
due to foreground cleaning violates the assumption that the noise in 
temperature is vanishing, used to obtain Eqs. (\ref{eq:dnslike},\ref{eq:qtilde},\ref{eq:utilde}) from Eq. (\ref{eq:likelihood}). 

Two masks are considered: KQ85y7 for T and P06 for (Q, U). Monopole and dipole have been subtracted from the 
observed ILC map through the HealPix routine {\sc remove-dipole} 
\citep{gorski}. The same data set has been used for the WMAP 7 yr power spectrum re-analysis by 
the Quadratic Maximum Likelihood (QML) estimator BolPol in \cite{Gruppuso2010} (similar data set for WMAP 5 yr data were 
previously used in \cite{Gruppuso2009,Paci:2010wp}). 
%and no differences have been found with respect to the WMAP public results.

\section{Cosmological Parameters Extraction}
\label{cosmopar}

We use \texttt{CosmoMC} \citep{cosmomc}
in order to compute the Bayesian probability distribution of model
parameters. The pivot scale of the primordial scalar and tensor power
spectra was set to $k_*=0.017$~Mpc$^{-1}$, as recommended by
\cite{Cortesetal}.  
We vary the physical baryon density $\Omega_{\rm b} h^2$, the physical cold dark matter density 
$\Omega_{\rm c}h^2$, the ratio of the sound horizon to the angular diameter distance at decoupling $\theta$, 
%the Hubble parameter $H_0=100 h \, {\rm km}\,{\rm s}^{-1}{\rm Mpc}^{-1}$ 
the reionisation optical depth $\tau$, the amplitude and spectral index of curvature perturbations $n_S$ and 
$\log_{10} [10^{10} A_s]$. We assume a flat universe, and so the cosmological constant for each
model is given by the combination $\Omega_{\Lambda} = 1-\Omega_{\rm
b}-\Omega_{\rm c}$. We set the CMB temperature $T_{\rm
CMB}=2.725$~K \citep{mather1999} and the primordial helium fraction to $y_{\rm He}=0.24$.
We assume three neutrinos with a negligible mass. 
In order to fit WMAP data, we use the lensed CMB and we follow the method implemented in 
\texttt{CosmoMC} consisting in varying a nuisance parameter $A_{\rm
SZ}$ which accounts for the unknown amplitude of the thermal SZ
contribution to the small-scale CMB data points assuming the model of
\cite{KoSe}.  We use \texttt{CAMB} \citep{camb} with accuracy setting
of $1$. We sample the posterior using the
Metropolis-Hastings algorithm \citep{Hastings} at a temperature
$T=1$, generating four 
parallel chains and imposing a conservative Gelman-Rubin convergence
criterion \citep{GelmanRubin} of $R-1 < 0.005$.

With the settings specified above we extract cosmological parameters with 
the WMAP likelihood code (version v4p1) available at http://lambda.gsfc.nasa.gov/ as benchmarks. We prefer 
to not quote the estimates for the cosmological parameters performed by the WMAP team since the conventions and 
the CAMB version might differ from 
those used in \cite{Larson:2010gs,Komatsu:2010fb}.

We then extract cosmological parameters by substituting 
the WMAP low-$\ell$ likelihood approach with BoPix. In doing this we implicitly use the WMAP inputs in polarization at $N_{\rm side}=16$
as described in Section III and not those contained in the WMAP likelihood routine publicly available. Since temperature and polarization 
are treated at the same resolution by BoPix, we include the WMAP high $\ell$ likelihood starting at $\ell =31$ both in temperature and temperature-polarization 
cross-correlation when using BoPix, unless otherwise stated. Unless otherwise stated, 
in BoPix we vary the $C_\ell$ up to $\ell=30$ and we use the publicly available file {\sl test\_cls\_v4.dat}
as a fiducial power spectrum to complete the full covariance at low resolution from $\ell=31$ to $\ell=64$, as done for temperature only by the 
WMAP pixel likelihood.

\begin{table*}
\centering
%\begin{minipage}{126mm}
%\caption{Cosmological parameters from WMAP 7 year: Correct} % title of Table
\centering % used for centering table
\begin{tabular}{c c c c c c c c} % centered columns (6 columns)
\hline\hline %inserts double horizontal lines
%Parameter & WMAP 7 & $l_{\rm trans} = 30$ &  $l_{\rm trans}=24$ & $l_{\rm trans}=36$ & Different &
%Same mask (WRONG) \\ & likelihood & & & & Fiducial  & in $T, P$ \\ [0.5ex] % inserts table
Parameter & WMAP 7 & WMAP 7 & $\ell_T = \ell_P = 30$ &  $\ell_T=\ell_P=24$ & $\ell_T=\ell_P=36$ & $\ell_T=30$ & Different 
\\ & likelihood & likelihood & & & &$\ell_P=23$ & Fiducial  \\ 
\\ & (Pixel) & (Gibbs) & & & & & \\
[0.5ex] % inserts table
\hline % inserts single horizontal line
100 $\Omega_b h^2$ & $2.250 \pm 0.056$& $2.252^{+0.057}_{-0.056}$ & 
$2.213 \pm 0.055$ & $2.215 \pm 0.055$ & 
$2.224^{+0.057}_{-0.058}$
& $2.213^{+0.055}_{-0.054}$ & $2.212 \pm 0.058 $ \\
%&  $2.230^{+0.056}_{-0.055}$ \\ % inserting body of the table
$\Omega_c h^2$ & $0.1114^{+0.0054}_{-0.0053}$ & $0.1114 \pm 0.0055$ & $0.1145^{+0.0055}_{-0.0056}$ &
$0.1142 \pm 0.0055$ & $0.1152^{+0.0055}_{-0.0056}$ & $0.1145^{+0.0056}_{-0.0057}$
& $0.1144 \pm 0.0056$ \\
%& $0.1119\pm0.0054$ \\
%$\theta $ & $1.0391^{+0.0027}_{-0.0026}$ & $1.0386 \pm 0.0027$ & $1.0387^{+0.0027}_{-0.0026}$ & $1.0384^{+0.0027}_{-0.0026}$ 
%& $1.0386^{+0.0027}_{-0.0026}$   & $1.0390 \pm 0.0026$ \\ % [1ex] adds vertical space
$\tau$ & $0.089 \pm 0.015$ & $0.089^{+0.014}_{-0.015}$ & 
$0.085^{+0.015}_{-0.014}$ &
$0.085^{+0.015}_{-0.014}$ & $0.085^{+0.014}_{-0.015}$ & $0.085^{+0.015}_{-0.014}$ 
& $0.085 \pm 0.015$ \\
%& $0.0910^{+0.0060}_{-0.0074}$ \\
$n_s$ & $0.968^{+0.014}_{-0.013}$& $0.969^{+0.013}_{-0.014}$ 
& $0.956 \pm 0.014$ & $0.957^{+0.014}_{-0.013}$ & $0.954^{+0.013}_{-0.014}$ &
$0.955 \pm ^{+0.014}_{-0.013}$ &
$0.956^{+0.014}_{-0.013}$ \\
%& $0.962^{+0.014}_{-0.013}$ \\
${\rm log} [10^{10} A_s]$ & $3.116^{+0.033}_{-0.032}$ 
& $3.116 \pm 0.033$ & $3.130 \pm 0.033$
& $3.128 \pm 0.032$
& $3.133^{+0.032}_{-0.033}$ & 
$3.129 \pm 0.032$
& $3.128 \pm 0.033$ \\
%& $3.127^{+0.032}_{-0.033}$
%\\
\hline %inserts single line
$\Omega_M$ & $0.270^{+0.027}_{-0.028}$ & $0.269 \pm 0.028$ & $0.289^{+0.031}_{-0.030}$ & $0.288^{+0.030}_{-0.031}$ 
& $0.294 \pm 0.031$ & $0.290 \pm 0.031$ 
& $0.289 \pm 0.030$ \\
$H_0$ & $70.7^{+2.4}_{-2.5}$ & $70.7^{+2.5}_{-2.4}$ & $68.9^{+2.3}_{-2.4}$ & $69.1 \pm 2.4$
& $68.5 \pm 2.4$ & $68.8 \pm 2.4$
& $68.9^{+2.3}_{-2.4}$ \\
%& $70.2 \pm 2.5$ \\
$\sigma_8$ & $0.811 \pm 0.029$ & $0.811 \pm 0.029$ & $0.820 \pm 0.029$
& $0.819^{+0.028}_{-0.029}$ & $0.822^{+0.030}_{-0.028}$ & $0.819 \pm 0.029$
& $0.820^{+0.029}_{-0.030}$ \\
%& $0.814^{+0.029}_{-0.030}$ \\
\end{tabular}
\label{marginalized} % is used to refer this table in the text
\caption{Mean parameter values and bounds of the central 68\%-credible intervals 
for the cosmological parameters estimated by the WMAP 7 year full likelihood (second and third column) 
and by the BoPix plus WMAP 7 year high $\ell$ likelihood for different transition 
multipoles $\ell_T=\ell_P$ (fourth, fifth and sixth column), for $\ell_T \ne \ell_P$  
and different fiducial 
theoretical power spectrum to complete the signal covariance matrix in BoPix (last column).  
Below the thick line analogous mean values and bounds are presented for derived parameters.}
%\end{minipage}
\end{table*}

%No appreciable differences have been noticed by constructing the signal covariance matrix up to 
%$3 {\rm Nside}$ instead up to 
%$4 {\rm Nside}$. This can be understood since this different prescription in constructing the signal 
%covariance matrix is damped by the Gaussian smoothing in intensity and is much below the noise in polarization.

\begin{figure}
\includegraphics[width=9cm,height=5cm,angle=0]{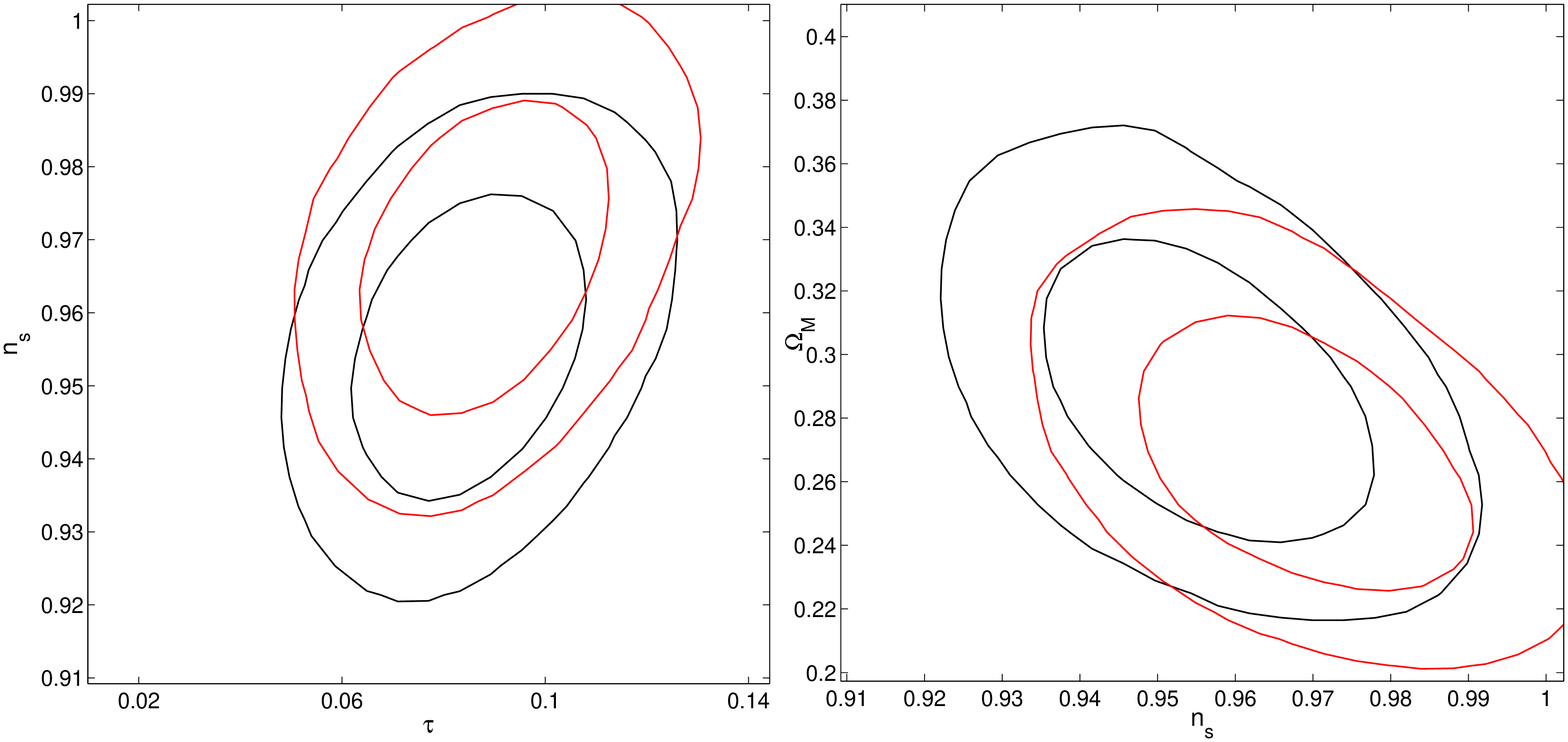}
\label{figurauno}
\caption{Marginalized 68\% and 95\%-credible contours for $(\tau \,, n_s)$ (left panel) and $(n_s \,, \Omega_M)$ (right panel)
as estimated by the WMAP 7 year full likelihood (red lines)
and by the BoPix plus WMAP 7 year high $\ell$ likelihood (black lines).}
\end{figure}

%{\bf Discussions}

We find small differences in the estimate of the cosmological parameters by substituting BoPix to the WMAP low-$\ell$ likelihood, as 
reported in Table I. 
\footnote{Note that the small differences of our results with the full WMAP 7 year likelihood with respect
to the results reported by \cite{Larson:2010gs} or \cite{Komatsu:2010fb} might be ascribed to
the different version of RECFAST used, different tools for extracting
cosmological parameters or different conventions, such as the pivot scale $k_*$.} 
The main difference between the estimate of the cosmological parameters derived by our alternative low-$\ell$ likelihood code and the one 
obtained with the WMAP approach is in the spectral index $n_s$: we obtain a value for $n_s$ which is 
0.86$\sigma$ lower than the WMAP one. This change would lead to quantitative differences in the evidence against the 
Harrison-Zeldovich of the WMAP 7 yr data.
However, also the other directly sampled cosmological parameters differ from the WMAP estimate in about $0.5 \sigma$, pointing towards 
values higher for the physical CDM abundance $\Omega_c h^2$ and the amplitude of scalar perturbations $A_S$ and smaller for the baryon physical 
content $\Omega_b h^2$ and optical depth $\tau$. As a derived parameters, we have a higher value 
for the matter content $\Omega_M$ and $\sigma_8$, smaller for the present Hubble rate $H_0$. We show more details about these different estimates in the 
two-dimensional plots of Fig. 1.
These differences seems robust to the change in the multipole transition to the high likelihood approximation and to the change of the 
fiducial model to complete the covariance at low resolution. 
Special mention should be made for the case in which we do not consider $\ell_T=\ell_P$, but we 
adopt the same $\ell_T=30$ and $\ell_P=23$ adopted by the WMAP team, but with BoPix 
for low resolution: the differences with respect to the estimates by the full WMAP yr likelihood are slightly 
smaller than in the case of $\ell_T=\ell_P=30$, as can be seen in Table I.
%Beyond what reported in Table I, slightly differences with respect to the use of the full WMAP 7 yr likelihood 
%are noticed when $C_\ell^{TE} \,, C_\ell^{EE}$ are varied only up to $\ell=23$: 
%with the last settings we get $n_s = 0.958^{+0.015}_{-0.014}$ and $\tau=0.087 \pm 0.015$, for instance. 
This means that differences we find are not fully due to the different threshold multipoles for polarization adopted in the 
two low-$\ell$ likelihood approaches.
No appreciable differences are noticed by constructing the signal covariance matrix up to
$3 N_{\rm side}$ instead up to
$4 N_{\rm side}$. This can be understood since this different prescription in constructing the signal
covariance matrix is damped by the Gaussian smoothing in intensity and is much below the noise in polarization. 

We have performed a further test excluding $A_{\rm SZ}$, just for code comparison. 
We find a smaller discrepancy between the estimates for the cosmological parameters and the best-fits 
from the two likelihood approaches when the nuisance parameter $A_{\rm SZ}$ is omitted (i.e. fixed to zero). This additional foreground 
parameter $A_{\rm SZ}$ is not well constrained by WMAP, but it contributes  
to the shape of the final likelihood and to the marginalized values of the parameters (shifting slightly the value of $n_s$, for instance).
We have checked that the different realizations of the $\mu$K rms noise added to the ILC temperature map in the WMAP and BoPix likelihood lead to 
much smaller differences than those reported.

Most of these small differences reported in the estimate of the cosmological parameters interfere 
destructively because of the cosmic confusion 
\citep{EfstathiouandBond} and the  
best-fits $C_\ell$ from the two likelihood analysis agree very well.
We present the CMB bestfit $C_\ell$ in temperature and lensing (the latter not entering in the 
likelihood evaluation) obtained by BoPix in combination with the WMAP 7 high-$\ell$ likelihood 
in comparison with those obtained by the full WMAP 7 likelihood in Fig. 2. 
The difference in the best-fit $C_\ell$ in temperature is consistent 
with the different central values for the cosmological parameters displayed in Table I. 
Note how the relative difference in the lensing is slightly larger than the one in temperature and does not 
decrease at high multipoles. Differences in polarization and temperature-polarization cross-correlation are 
smaller than the ones shown here.
We have checked that the best-fit $C_\ell$ obtained in this work by the full WMAP 7 likelihood has 
$\Delta (-2 \log {\cal L}_{\rm WMAP})=-7.42$ with 
respect to the 
reference WMAP 7 {\sl test\_cls\_v4.dat}; the best-fit $C_\ell$ obtained in this work by BoPix in combination with 
the high-$\ell$ WMAP 7 likelihood provides 
a better fit, with $\Delta (-2 \log {\cal L}_{\rm WMAP})=-7.75$ with respect to the reference 
WMAP 7 {\sl test\_cls\_v4.dat}.

\begin{figure}
\includegraphics[width=9cm,height=5cm,angle=0]{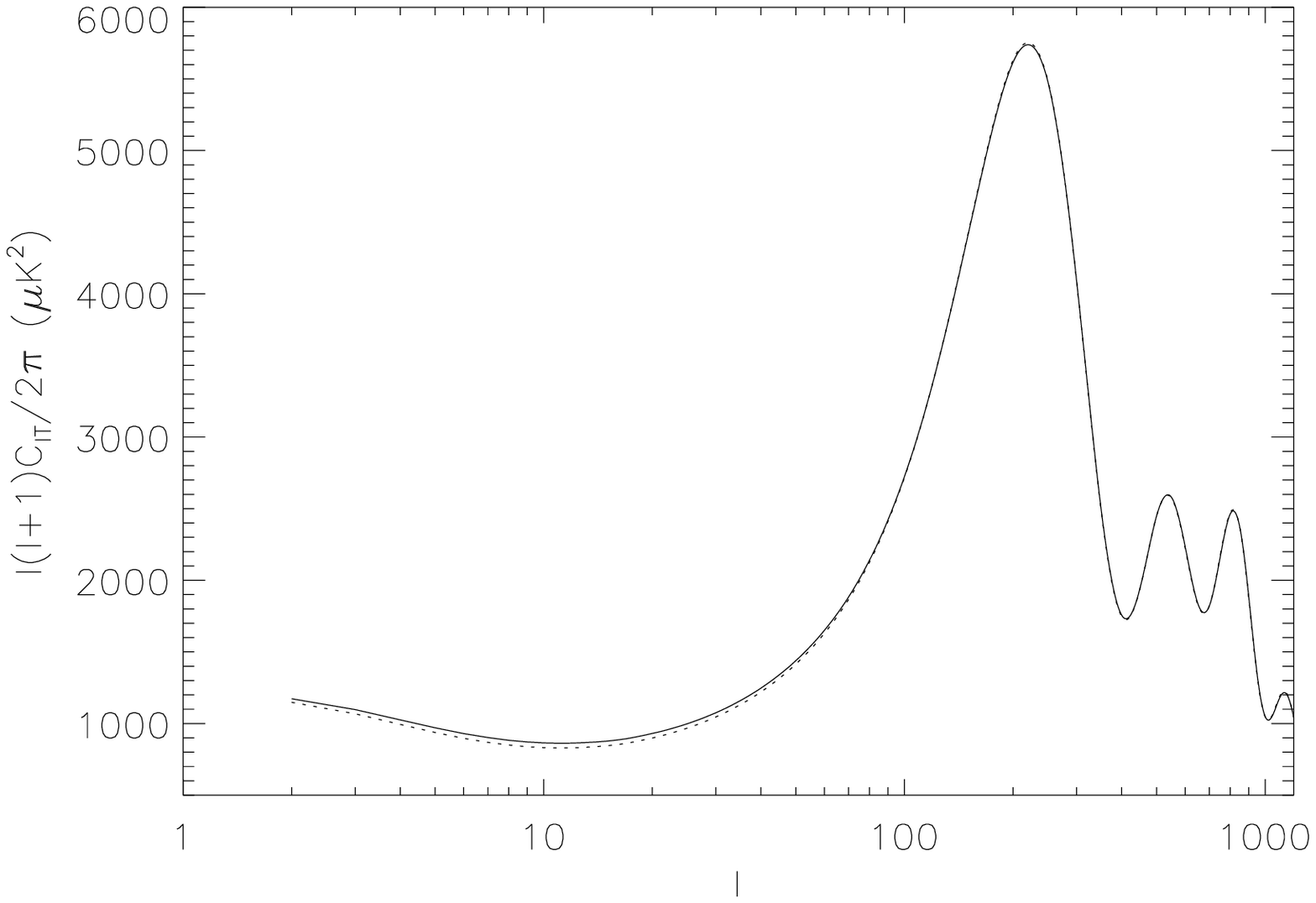}
\includegraphics[width=9cm,height=5cm,angle=0]{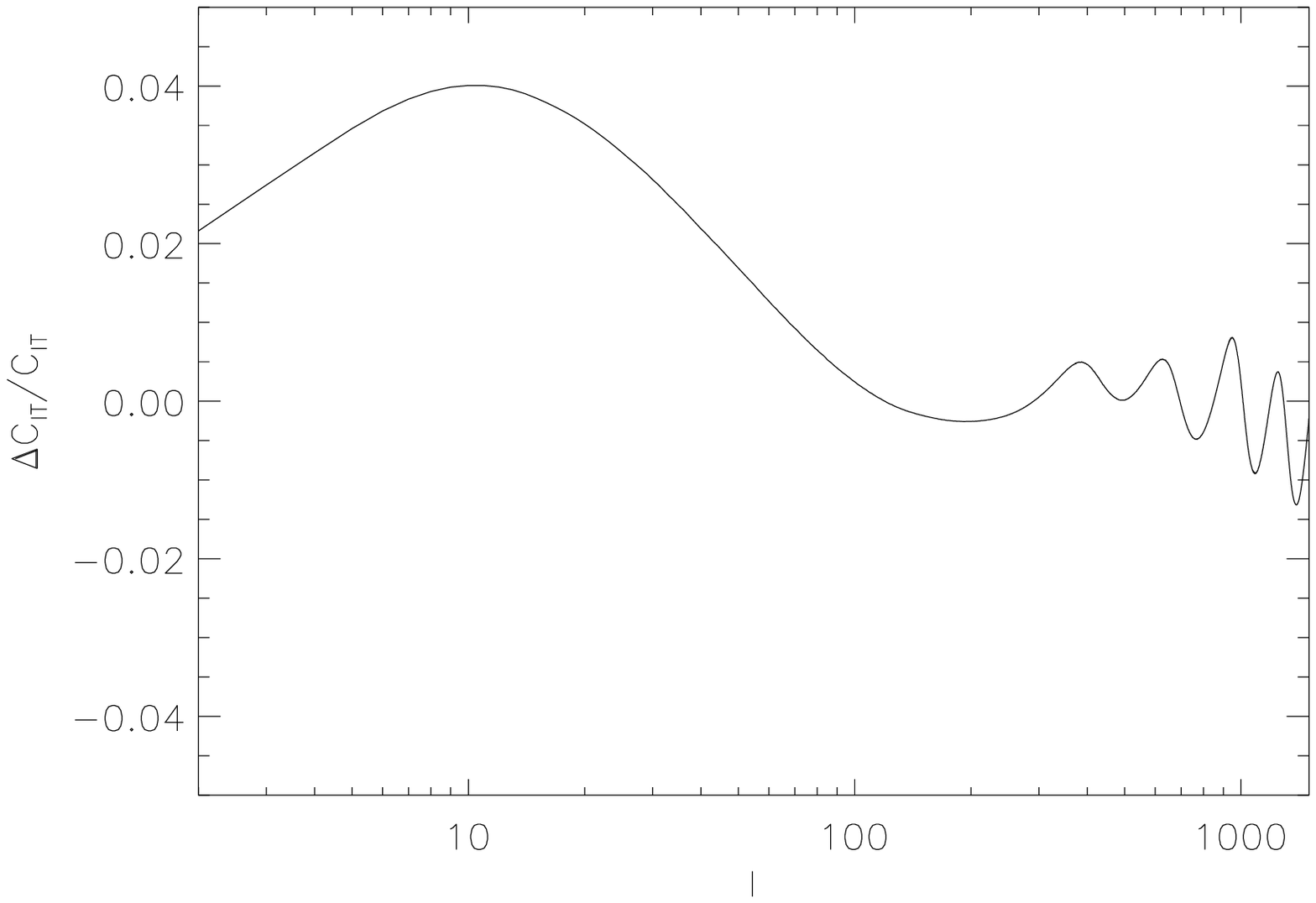}
\includegraphics[width=9cm,height=5cm,angle=0]{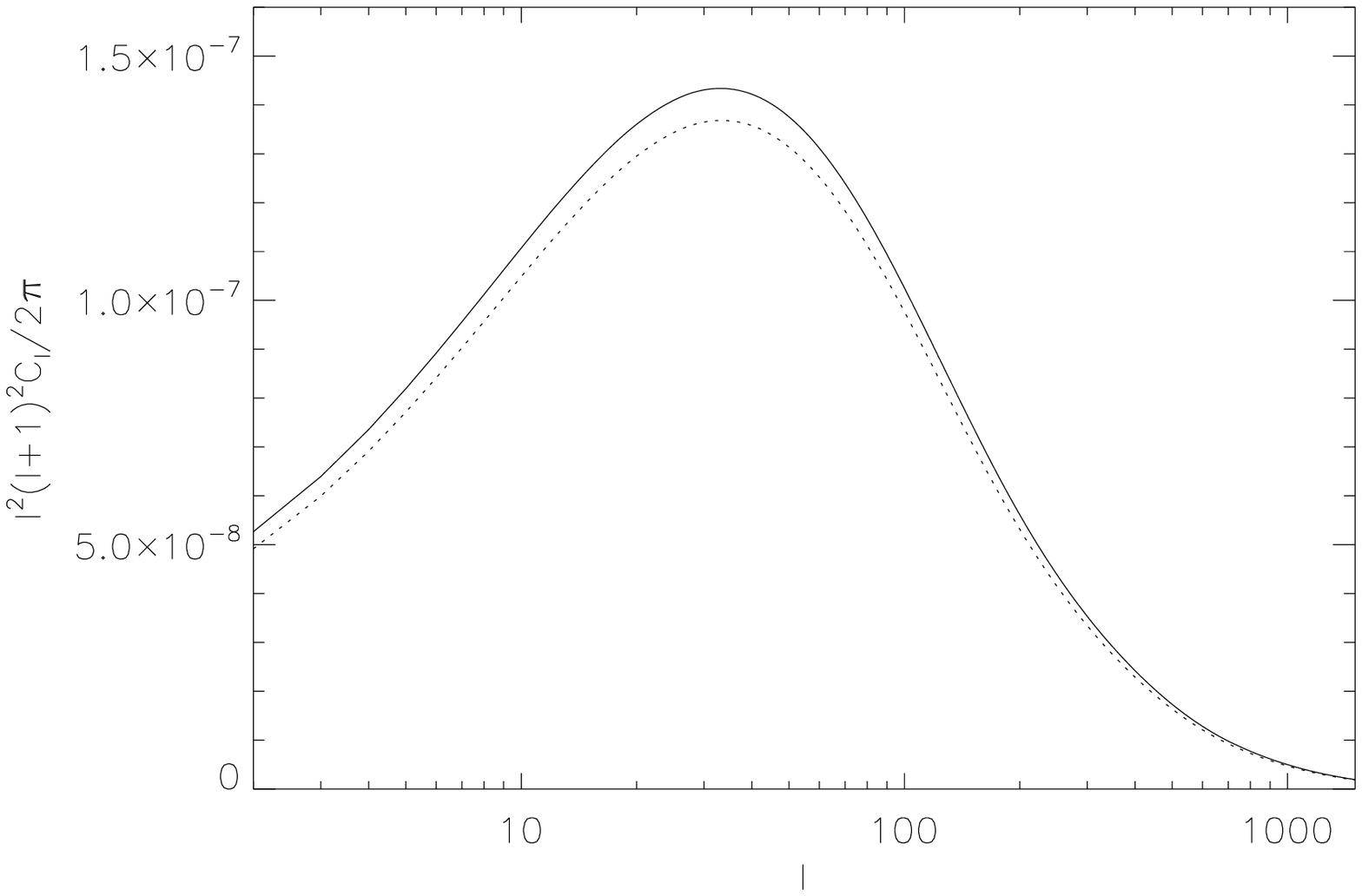}
\includegraphics[width=9cm,height=5cm,angle=0]{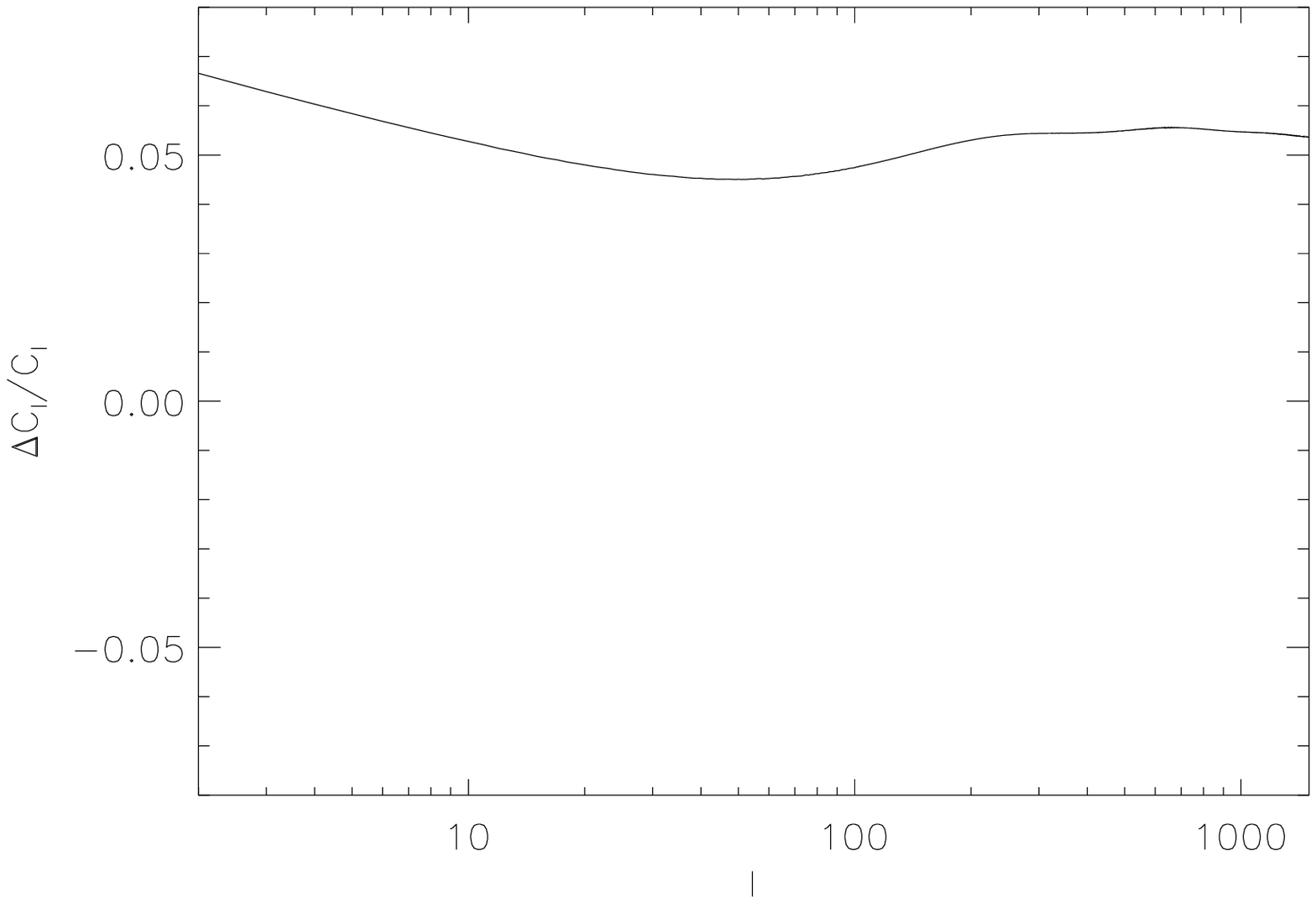}
\label{figurabestfit}
\caption{Comparison of the best-fit $\ell(\ell+1)C_\ell^{TT}/(2 \pi)$ and 
$\ell^2 (\ell+1)^2 C_\ell^{\phi \phi}/(2 \pi)$
obtained by BoPix in combination with the WMAP 7 high-$\ell$ 
likelihood (solid) vs. the WMAP 7 full likelihood (dashed) is shown in the first and third panel from above. 
%The temperature (temperature-polarization cross-correlation) $C_\ell$ is shown in the top (middle) panel. 
To make the difference more visible, the relative difference between the $C_\ell$ bestfits in temperature 
and lensing potential are shown in the second and fourth panels, respectively. Note that 
the differences are well within the cosmic variance.}
\end{figure}

%Since the pixel approach and the Gibbs approach contained in the WMAP likelihood package give quite consistent results, 
%the difference we find by using BoPix is more likely due to the asymmetric treatment of $T$ and $(Q,U)$ adopted in WMAP low-$\ell$ likelihood. 
We have then tested BoPix against the WMAP likelihood within the same range of multipole, i.e. up to $\ell=30$: 
BoPix has been run on the low-resolution WMAP 7 yr $N_{\rm side}=16$ 
products varying $C^{TT}_\ell$, $C^{EE}_\ell$, $C^{TE}_\ell$ up to $\ell=30$ and compared 
to the likelihood obtained by the WMAP 7 yr pixel based 
routine plus 
the high-$\ell$ likelihood value for TE from $\ell=24$ to $\ell=30$. In this way we subtract the {\em same} high-$\ell$ likelihood information 
from hybrid runs 
presented in 
Table I. By assuming $\Omega_b h^2 = 0.02246$, $\Omega_c h^2 = 0.1117$ and sound horizon $\theta = 1.03965$,  
we obtain results quite consistent with the hybrid ones: a slight smaller value 
in the estimate of $\tau$ and $n_S$ and a larger one for $A_S$, as shown in Fig. 3. 

\begin{figure}
\includegraphics[width=9cm,height=5cm,angle=0]{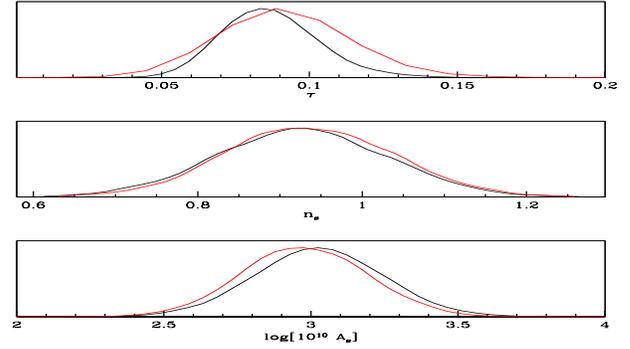}
\label{figuralowl}
\caption{Marginalized one-dimensional probabilities for $\tau$, $n_s$ and ${\rm log} [10^{10} A_s]$
as estimated by the WMAP 7 year full likelihood (red lines)
and by the BoPix plus WMAP 7 year high $\ell$ likelihood (black lines). See text for further details.}
\end{figure}

As already mentioned, one important aspect of the WMAP 7 year low-$\ell$ 
likelihood is to use two different resolution for temperature and polarization;  
the polarization information at HEALPIX resolution $N_{\rm side}=8$ 
is used up to the Nyquist multipole, i.e. $\ell_P=23$. 
We run the two low-$\ell$ likelihoods with $\ell_T=\ell_P=16$ to make sure that the 
differences are not due mainly to a mismatch in the polarization 
data sets. As reported in Table 2, the differences in the estimates of the parameters 
decrease, as expected, but do not disappear.

\begin{table}%[h]
%\caption{Cosmological parameters from WMAP 7 year including the tensor-to-scalar ratio} % title of Table
\centering % used for centering table
\begin{tabular}{c c c} % centered columns (3 columns)
\hline\hline %inserts double horizontal lines
Parameter & WMAP 7 & BoPix plus \\
& likelihood &  WMAP 7 high $\ell$ likelihood\\
[0.5ex] % inserts table
%heading
\hline % inserts single horizontal line
100 $\Omega_b h^2$ & $2.246 \pm 0.057$ & $2.231^{+0.0057}_{-0.0058}$ \\ % inserting body of the table
$\Omega_c h^2$ & $0.1119 \pm 0.0055$ & $0.1113 \pm 0.0067$ \\
%$\theta $ & $1.0403^{+0.0028}_{-0.0029}$ & $1.0398^{+0.0029}_{-0.0030}$ \\ [1ex] % [1ex] adds vertical space
$\tau$ & $0.088 \pm 0.015$ & $0.087 \pm 0.015$ \\
$n_s$ & $0.967^{+0.013}_{-0.014}$ & $0.962^{+0.014}_{-0.015}$ \\
${\rm log} [10^{10} A_s]$ & $3.118 \pm 0.033$ & $3.117\pm 0.033$ \\
\hline
$\Omega_M$ & $0.273 \pm 0.029$ & $0.271^{+0.030}_{-0.029}$ \\
$H_0$ & $70.4^{+2.5}_{-2.4}$ & $70.4 \pm 2.6$ \\
$\sigma_8$ & $0.812 \pm 0.030$ & $0.807 \pm 0.030$ \\
\end{tabular}
\label{marginalized_r} % is used to refer this table in the text
\caption{Mean parameter values and bounds of the central 68\%-credible intervals
for the cosmological parameters with a transition in the hybrid likelihood at $\ell=16$. 
The results of the WMAP 7 year full likelihood 
(BoPix plus WMAP 7 year high $\ell$ likelihood) are reported in the left (right) column.
Below the thick line analogous mean values and bounds are
presented for derived parameters.}
\end{table}

Another important difference between BoPix and the WMAP 7 yr likelihood routine is the 
treatment of monopole and dipole for the temperature map. In the 
ILC temperature map with the additional noise of 1 $\mu K$ rms used in BoPix, 
the monopole and dipole in the masked sky are removed; no monopole and dipole terms are 
considered in the construction of the covariance matrix.
The WMAP 7 yr temperature pixel routine instead does not subtract the monopole and dipole in the masked  
sky; in the observed sky with the KQ85y7 mask,
the ILC temperature map has an offset of -0.07 $\mu K$ and a dipole 
$C_1=4.6 \, \mu K^2$. 
To take into account monopole and dipole residuals, 
the full sky signal covariance matrix is modified according to 
\cite{Slosar04}:
\be
S (\theta) \rightarrow S (\theta) + \lambda \left ( \frac{P_0}{4 \pi} 
+ \frac{3}{4 \pi} P_1 \right)  
\label{C0mod}
\ee
where $P_0(\cos \theta)=1$ and $P_1(\cos \theta)=\cos \theta$ 
are the Legendre polynomials associated to monopole and dipole, respectively. 
The fixed amplitude of the monopole and dipole terms is taken to be equal to 
the quadrupole of the fiducial $\Lambda$CDM model, i.e. $\lambda=1262 \mu K^2$. 
%In the observed sky with the KQ85y7 mask, 
%the ILC temperature map has an offset of -0.07 $\mu K$ and a dipole of $4.6 \, \mu K^2$.
%Although called monopole and dipole marginalization, the 
The subtraction of monopole and dipole in the masked ILC map has a little impact on the estimate of cosmological 
parameters. Cosmological parameters instead have a 
strong dependence on the amplitude $\lambda$ of the monopole and dipole terms which contribute 
to the signal covariance matrix, as shown in Fig. 4. 
The results obtained by 
subtracting monopole and dipole in the ILC temperature map used by the WMAP 7 yr temperature 
pixel likelihood 
routine and setting $\lambda=0$ in the construction of the temperature 
covariance matrix do not match with those obtained by BoPix, as shown in Fig. 4. Viceversa, by tuning 
the amplitude of the monopole and dipole term to $0.17 \mu K^2$ the results of the 
WMAP 7 yr likelihood routine agrees with those by BoPix. We conclude that part, but not all, of the discrepancy between 
BoPix and WMAP 7 yr likelihood is due to the monopole and dipole marginalization in Eq. \ref{C0mod}. 
%The explanation 
%is simple: in the observed sky, the monopole and dipole terms constitute an additional contribution on large 
%angular scales, which is compensated by a cosmological model with a less power on large scales, i.e. 
%a higher value for $n_s$, with respect to the case in which no monopole and dipole terms are considered.

%C0

\begin{figure}
\includegraphics[width=9cm,height=5cm,angle=0]{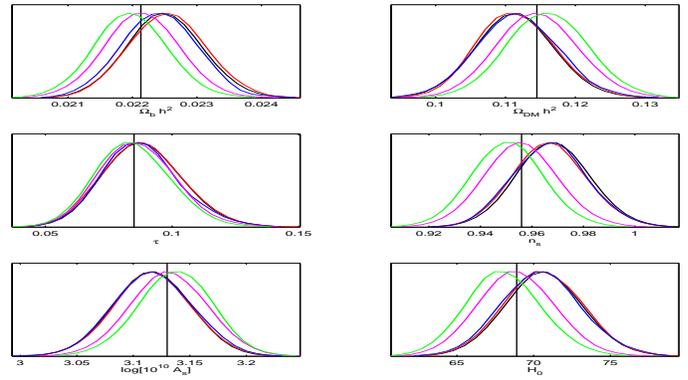}
\label{figuraC0}
\caption{Marginalized one-dimensional probabilities for cosmological parameters  
as estimated by the WMAP 7 year full likelihood for $\lambda=1262 \mu K^2$ (black line), 
$\lambda=12.62 \mu K^2$ (red line), $\lambda=1.262 \mu K^2$ (blu line), 
$\lambda = 0.168 \mu K^2$ (purple line). The green line is obtained with the WMAP 
by removing monopole and dipole in the masked sky and setting $\lambda=0$. 
The black vertical lines are 
the mean values obtained by 
BoPix in combination with the 
WMAP 7 yr high $\ell$ likelihood listed in the fourth column of 
Table 1, which agree with the central values of the posteriors in purple.}
\end{figure}

%We end this section in noting that the value of $n_s$ we find is consistent with the value found

\section{Other extended cosmological models}

We now consider few cosmological models beyond the $\Lambda$CDM model which can be constrained by WMAP 7 year data only.
We consider only the baseline $l_{\rm trans} = 30$ and 
all the other settings consistently with the previous section, unless otherwise stated.

{\em Gravitational Waves.} 

\noindent
We consider all inflationary models which can be described by the
primordial perturbation parameters consisting of the scalar amplitude and
spectral index $(A_{\rm S}, n_{\rm S})$, and the tensor-to-scalar
ratio $r$. 
%(both defined at the pivot scale $k_*=0.017$~Mpc$^{-1}$ consistently with the previous section.).
In canonical single-field inflation, in the slow-roll limit, the tensor
spectrum shape is not independent of the scalar one. We will consider
a tensor spectrum with a tilt $n_{\rm T}=-r/8$, as predicted for
canonical single-field inflation at first-order in slow-roll.

Our marginalised 68\%-credible interval for the scalar spectral index is given by
$n_{\rm S}=0.977^{+0.020}_{-0.021}$, 
%(to be compared with the result
%of the previous section, $n_{\rm S}=0.955 \pm {0.014}$), 
half a sigma redder than the result we obtain 
by the full WMAP 7 year likelihood $0.987 \pm 0.020$.

At 95\% confidence level, our result for the tensor-to-scalar
ratio is $r < 0.36$, fully consistent with the result we obtain from the full WMAP 7 year likelihood, i.e. 
$r<0.34$. Let us note that, differently from the WMAP low-$\ell$ likelihood code, 
BoPix include BB polarization in the construction of the covariance at low resolution.
Estimates of the cosmological parameters including tensor modes are compared in Table 3.
The differences in the $(n_{\rm S}, r)$ are shown in Fig. 5 and are mainly due to a shift of the constraints at 
smaller values for $n_{\rm S}$, as occurs for the standard $\Lambda$CDM model discussed in the previous section. 
Theoretical predictions of few popular inflationary models (including reheating uncertainties where appropriate)
are displayed. One of the phenomenological differences from the different constraints would be a minor 
tension for a massless self-interacting inflaton model with WMAP 7 year data {\em only} 
(see \cite{Komatsu:2010fb,Finellietal2010} as examples for an higher tension of the 
$\lambda \phi^4$ potential with observations 
when additional cosmological data sets are added to WMAP). 

\begin{table}%[h]
%\caption{Cosmological parameters from WMAP 7 year including the tensor-to-scalar ratio} % title of Table
\centering % used for centering table
\begin{tabular}{c c c} % centered columns (3 columns)
\hline\hline %inserts double horizontal lines
Parameter & WMAP 7 & BoPix plus \\
& likelihood &  WMAP 7 high $\ell$ likelihood\\
[0.5ex] % inserts table
%heading
\hline % inserts single horizontal line
100 $\Omega_b h^2$ & $2.307^{+0.071}_{-0.072}$ & $2.270 \pm 0.073$ \\ % inserting body of the table
$\Omega_c h^2$ & $0.1073 \pm 0.0063$ & $0.1099^{+0.0067}_{-0.0066}$ \\
%$\theta $ & $1.0403^{+0.0028}_{-0.0029}$ & $1.0398^{+0.0029}_{-0.0030}$ \\ [1ex] % [1ex] adds vertical space
$\tau$ & $0.091^{+0.015}_{-0.014}$ & $0.087^{+0.015}_{-0.014}$ \\
$n_s$ & $0.987 \pm 0.020$ & $0.977^{+0.020}_{-0.021}$ \\
${\rm log} [10^{10} A_s]$ & $3.093 \pm 0.038$
& $3.102\pm 0.039$ \\
$r$ & $< 0.34$ & $< 0.36$ \\ \hline %inserts single line
$\Omega_M$ & $0.246^{+0.031}_{-0.032}$ & $0.262^{+0.035}_{-0.036}$ \\
$H_0$ & $73.2 \pm 3.2$ & $71.6^{+3.2}_{-3.3}$ \\
$\sigma_8$ & $0.797 \pm 0.033$ & $0.805 \pm 0.033$ \\
\end{tabular}
\label{marginalized_r} % is used to refer this table in the text
\caption{Mean parameter values and bounds of the central 68\%-credible intervals
for the cosmological parameters including the tensor-to-scalar ratio estimated by the WMAP 7 year full likelihood 
(left column) and by the BoPix plus WMAP 7 year high $\ell$ likelihood (right column). 
For the tensor-to-scalar ratio $r$ the 95\%-credible upper bound is quoted. Below the thick line analogous mean values and bounds are 
presented for derived parameters.}
\end{table}

\begin{figure}
\includegraphics[width=9cm,height=5cm,angle=0]{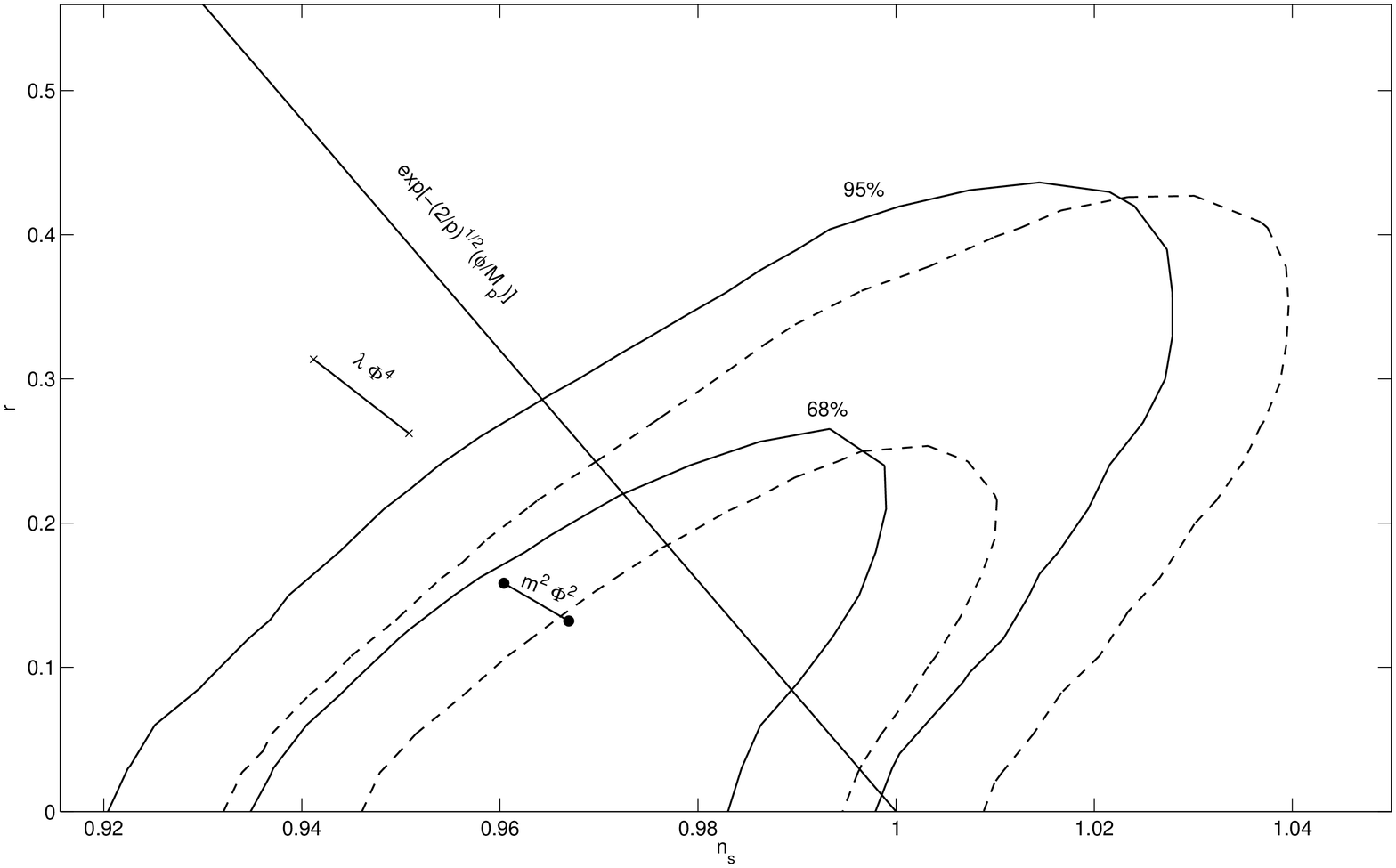}
\label{figuradue}
\caption{Marginalized 68\%\, 95\% contours for $(n_s \,, r)$
as estimated by the WMAP 7 year full likelihood (dashed lines)
and by the BoPix plus WMAP 7 year high $\ell$ likelihood (solid lines). 
Theoretical predictions of few popular inflationary models (including reheating uncertainties where appropriate) 
are displayed.}
\end{figure}

{\em Running of the scalar spectral index.}

\noindent
In this subsection we consider the variation of the scalar spectral index with wavelength, i.e. we allow $n_{\rm run}$
%$\alpha_{\rm S}$ 
to vary in the range $[-0.2,0.2]$. 
%Since one of the main changes in the estimate of cosmological parameters involved 
%$n_{\rm S}$, we expect that estimate of the running might be different as well. 
%Moreover, the running was indeed one of the main targets of independent likelihood 
%refinements after the WMAP first year release \cite{efstathiouetal}.  
Our marginalised 95\%-credible interval for the scalar spectral index is given by $-0.065 < n_{\rm run} < 0.042$,
%$-0.065 < \alpha_{\rm S} < 0.042$, 
which has to be compared with the result we obtain
by the full WMAP 7 year likelihood $-0.074 < n_{\rm run} < 0.030$. The results, shown in Table 4 
and Fig. 6,
are both consistent with the hypothesis of no wavelength dependence of the scalar spectral index. 

\begin{table}
%\caption{Cosmological parameters from WMAP 7 year including the running of the spectral index} % title of Table
\centering % used for centering table
\begin{tabular}{c c c} % centered columns (3 columns)
\hline\hline %inserts double horizontal lines
Parameter & WMAP 7 & BoPix plus \\
& likelihood &  WMAP 7 high $\ell$ likelihood\\
%Parameter & Full WMAP 7 likelihood & BoPix + WMAP 7 high $\ell$ likelihood\\
%& full likelihood & & & & & in $T,P$ \\
[0.5ex] % inserts table
%heading
\hline % inserts single horizontal line
100 $\Omega_b h^2$ & $2.198^{+0.074}_{-0.072}$ & $2.184 \pm 0.081$ \\ % inserting body of the table
$\Omega_c h^2$ & $0.1167 \pm 0.0082$ & $0.1175^{+0.0083}_{-0.0084}$ \\
%$\theta $ & $1.0386 \pm 0.0028$ & $1.0383 \pm 0.0027$ \\ % [1ex] adds vertical space
$\tau$ & $0.091^{+0.015}_{-0.016}$ & $0.087 \pm 0.015$ \\
$n_s$ & $0.961 \pm 0.016$ & $0.953^{+0.015}_{-0.016}$ \\
${\rm log} [10^{10} A_s]$ & $3.154 \pm 0.054$& $3.151^{+0.054}_{-0.055}$ \\
$n_{\rm run}$ & $-0.074 < n_{\rm run} < 0.030$ & $-0.065 < n_{\rm run} < 0.042$ \\
\hline %inserts single line
$\Omega_M$ & $0.303 \pm 0.049$ & $0.310^{+0.050}_{-0.051}$ \\
$H_0$ & $68.2^{+3.7}_{-3.6}$ & $67.5^{+3.8}_{-3.7}$ \\
$\sigma_8$ & $0.823^{+0.033}_{-0.032}$ & $0.826^{+0.033}_{-0.032}$ \\
\end{tabular}
\label{marginalized_running} % is used to refer this table in the text
\caption{Mean parameter values and bounds of the central 68\%-credible intervals
  for the cosmological parameters including the running of the scalar spectral index $n_{\rm run}$ 
estimated by the WMAP 7 year full likelihood
(left column) and by the BoPix plus WMAP 7 year high $\ell$ likelihood (right column).
For the running of the scalar spectral index $n_{\rm run}$ the 95\%-credible upper bound is quoted.}
\end{table}

\begin{figure}
\includegraphics[width=9cm,height=5cm,angle=0]{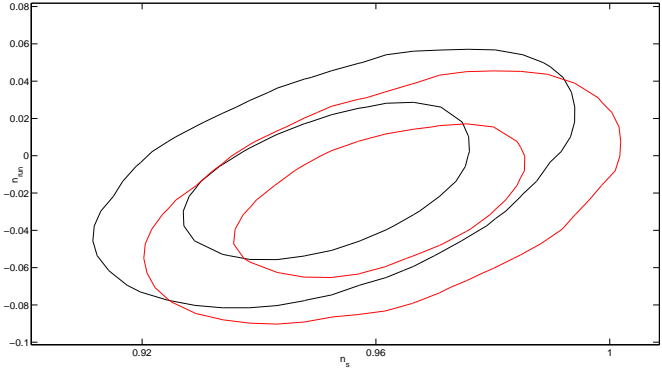}
\label{figuratre}
\caption{Marginalized 68\%\ and 95\%-credible contours for $(n_s \,, n_{\rm run})$ 
as estimated by the WMAP 7 year full likelihood (red lines)
and by the BoPix plus WMAP 7 year high $\ell$ likelihood (black lines).}
\end{figure}

{\em Neutrino Mass.}

In this subsection we constrain the total mass of neutrinos $\sum m_\nu = 94 \Omega_\nu h^2$ eV, 
allowing to vary the fraction of massive neutrino 
energy density relative to the total dark matter one $f_\nu =\Omega_\nu/\Omega_{\rm DM}$. 
At 95\% confidence level, our result for the fraction of massive neutrinos 
is $f_\nu < 0.113$, whereas we obtain $f_\nu < 0.094$ from the full WMAP 7 year likelihood. 
The resulting neutrino mass bound at 95\% confidence level is $\sum m_\nu < 1.4$ eV, compared to $1.1$ eV obtained 
from the full WMAP 7 year likelihood. The results are shown in Table 5 and Fig. 7.  

\begin{table}
%\caption{Cosmological parameters from WMAP 7 year including neutrino mass} % title of Table
\centering % used for centering table
\begin{tabular}{c c c} % centered columns (3 columns)
\hline\hline %inserts double horizontal lines
Parameter & WMAP 7 & BoPix plus \\
& likelihood &  WMAP 7 high $\ell$ likelihood\\
%Parameter & Full WMAP 7 likelihood & BoPix + WMAP 7 high $\ell$ likelihood\\
%& full likelihood & & & & & in $T,P$ \\
[0.5ex] % inserts table
%heading
\hline % inserts single horizontal line
100 $\Omega_b h^2$ & $2.219^{+0.062}_{-0.060}$ & $2.174 \pm 0.061$ \\ % inserting body of the table
$\Omega_c h^2$ & $0.1177^{+0.0071}_{-0.0073}$ & $0.1226^{+0.0081}_{-0.0080}$ \\
%$\theta $ & $1.0388 \pm 0.0027$ & $1.0383 \pm 0.0027$ \\ %[1ex] % [1ex] adds vertical space
$\tau$ & $0.087^{+0.014}_{-0.015}$ & $0.082 \pm 0.014$ \\
$n_s$ & $0.960 \pm 0.016$ & $0.945^{+0.016}_{-0.017}$ \\
${\rm log} [10^{10} A_s]$ & $3.120 \pm 0.032$
& $3.134 \pm 0.033$ \\
$f_\nu$ & $< 0.094$ & $<0.113$ \\ \hline %inserts single line
$\Omega_M$ & $0.329^{+0.057}_{-0.056}$  & $0.374^{+0.075}_{-0.072}$ \\
$H_0$ & $65.7^{+4.3}_{-4.2}$ & $62.8^{+4.6}_{-4.7}$ \\
$\sigma_8$ & $0.712^{+0.073}_{-0.74}$ & $0.695^{+0.087}_{-0.083}$ \\
$\sum m_\nu$ & $< 1.1$ eV & $< 1.4$ eV \\
\end{tabular}
\label{marginalized_nu} % is used to refer this table in the text
\caption{Mean parameter values and bounds of the central 68\%-credible intervals
  for the cosmological parameters including the total mass of the neutrinos
estimated by the WMAP 7 year full likelihood
(left column) and by the BoPix plus WMAP 7 year high $\ell$ likelihood (right column).
For the total mass of the neutrinos $\sum m_\nu$ the 95\%-credible upper bound is quoted.}
\end{table}

\begin{figure}
\includegraphics[width=9cm,height=5cm,angle=0]{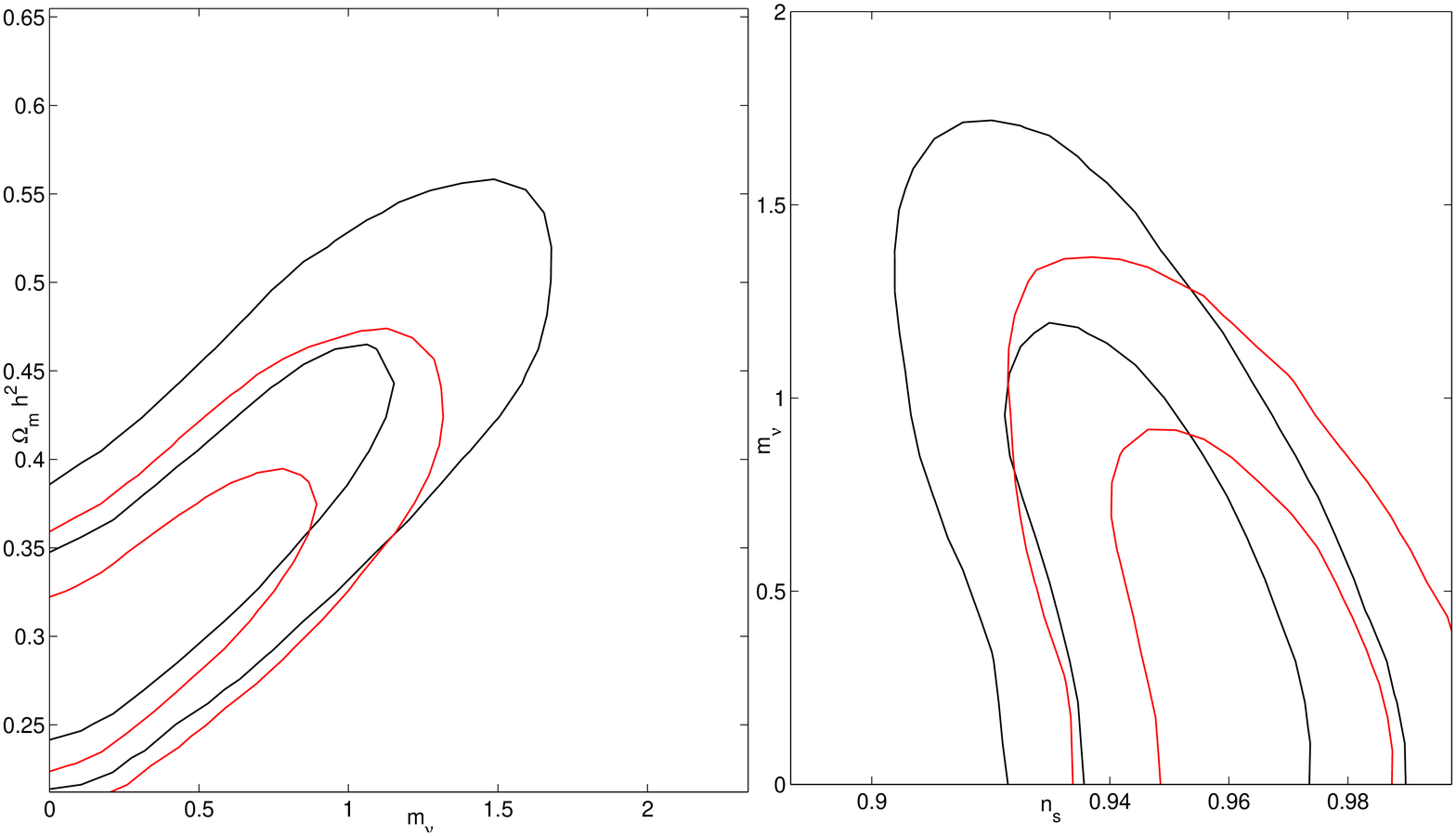}
\label{figuraquattro}
\caption{Marginalized 68\%\ and 95\%-credible contours for $(\sum_\nu m_\nu \,, \Omega_M h^2)$ (left panel) and 
$(n_s \,, \sum_\nu m_\nu)$ (right panel)             
as estimated by the WMAP 7 year full likelihood (red lines)
and by the BoPix plus WMAP 7 year high $\ell$ likelihood (black lines).}
\end{figure}

{\em Cosmological Birefringence.}

\noindent
Since one of the main differences between the WMAP low resolution likelihood code and BoPix is the treatment of the 
polarization sector, 
we now wish to analyze an extended cosmological model different 
from $\Lambda$CDM {\em only} in $(Q,U)$ and the relative cross-correlation with the temperature.
%We now analyze the constraints on cosmological birefringence obtained by our independent pixel-based likelihood code. 
Cosmological birefringence refers to a non-vanishing interaction $\propto \phi F_{\mu \nu} {\tilde F}^{\mu \nu}$
between photon and a cosmological evolving pseudo-scalar $\phi$, which would generate non-vanishing $TB$ and $EB$ 
correlations \citep{lueetal} through 
a rotation $\alpha$ of the polarization plane of CMB photons 
along their path from the last scattering surface to the observer. 
The resulting polarization and cross temperature-polarization spectra would encode the particular redshift dependence 
of the parity violation interaction \citep{liu,FinelliGalaverni2009}.
However, a phenomenological shortcut exists, commonly used in the literature and 
also adopted by the WMAP team, and consists to neglect the redshift dependence of $\alpha$ and 
simple predict the power spectra as \cite{lueetal}:
%By assuming $\bar{\theta} = \theta(\eta_0)$ where
%a very simple prediction for observed power spectra is indeed obtained \cite{Lue:1998mq}:
\begin{eqnarray}
%\label{C_ll_EE_constant}
C_\ell^{EE,obs}&=&C_\ell^{EE} \cos^2(2 \alpha)\,,\nonumber \\
%\label{C_ll_BB_constant}
C_\ell^{BB,obs}&=&C_\ell^{EE} \sin^2(2 \alpha)\,, \nonumber \\
\label{C_ll_EB_constant}
C_\ell^{EB,obs}&=&\frac{1}{2} C_\ell^{EE} \sin(4 \alpha)\,,\\
%\label{C_ll_TE_constant}
C_\ell^{TE,obs}&=&C_\ell^{TE} \cos(2 \alpha)\,, \nonumber \\
%\label{C_ll_TB_constant}
C_\ell^{TB,obs}&=&C_\ell^{TE} \sin(2 \alpha)\, \nonumber .
\end{eqnarray}
The above formulae are valid when the primordial B-mode polarization is negligible, which is assumed in this paper.

We have therefore sampled $\alpha$ in radiants with a flat prior $[-0.5 \, 0.5]$ plus the other six 
cosmological parameters of the $\Lambda$CDM model by inserting Eqs. (\ref{C_ll_EB_constant}).
As shown in Table \ref{marginalized_biri}, our marginalised 68\% (95\%)-credible interval for $\alpha$ is 
$\alpha=-1\,^{\circ}.3^{+0\,^{\circ}.6 \, +2\,^{\circ}.3}_{-0\,^{\circ}.7 \, -2\,^{\circ}.3}$ 
in agreement with the full WMAP 7 year likelihood result which we find 
$\alpha=-1\,^{\circ}.0^{+0\,^{\circ}.7 \, +2\,^{\circ}.4}_{-0\,^{\circ}.6 \, -2\,^{\circ}.3}$ 
%\footnote{Small differences with respect 
%to the results reported by \cite{Larson:2010gs} or \cite{Komatsu:2010fb} might be ascribed to 
%the different version of RECFAST used, different tools for extracting 
%cosmological parameters or different conventions, such as the pivot scale $k_*$.}. 
Either the result using BoPix or the one based on the full 
WMAP 7 year likelihood are consistent with vanishing cosmological birefringence at 95\% CL 
just by assuming the statistical uncertainty, and the 
agreement increases by using the systematic uncertainty, which is estimated as $1\,^{\circ}.4$ 
by the WMAP team \cite{Komatsu:2010fb}.

Since the weight of the high-$\ell$ TB likelihood plays a relevant role in these constraints we have also considered the 
case in which this is not taken into account. Such setting which emphasizes the role of polarization on large angular 
scales would be relevant to show clearly the potential differences between BoPix and the WMAP pixel likelihood code.
On using only low resolution products to constrain cosmological birefringence, 
by using BoPix on $N_{\rm side} = 16$ resolution $Q \,, U$ maps and matrices 
we obtain $\alpha=-4\,^{\circ}.2^{+1\,^{\circ}.9 \, +10\,^{\circ}.2}_{-3\,^{\circ}.1 \, -7\,^{\circ}.5}$, 
still in agreement with the values we find by the WMAP 7 likelihood on $N_{\rm side} = 8$ resolution 
$Q \,, U$ maps and matrices $\alpha=-0\,^{\circ}.2^{+3\,^{\circ}.6 \, +10\,^{\circ}.0}_{-3\,^{\circ}.6 \, -9\,^{\circ}.9}$ 
%\footnote{Small differences with respect
%to the results reported by \cite{Larson:2010gs} or \cite{Komatsu:2010fb} might be ascribed to
%the different version of RECFAST used, different tools for extracting
%cosmological parameters or different conventions, such as the pivot scale $k_*$.}. 
Although with larger uncertainties, our results agree with 
vanishing cosmological birefringence at 95\% CL, without invoking 
systematic uncertainties. Note also that our result agrees with 
the analysis on large angular scales by \cite{Gruppuso2011}, 
where much tighter constraints are given probably because all the cosmological 
parameters except $\alpha$ are kept fixed.

The full posterior likelihood for $\alpha$ and its two dimensional contour in combination 
with the optical depth $\tau$ are shown in 
Fig. 8, which shows that no degeneracy 
between $\tau$ and $\alpha$ is observed in WMAP 7 yr data. 
Note that the slight preference at 68\% CL for negative values of $\alpha$ 
when using only BoPix on low resolution products is consistent with the WMAP 7 yr TB and EB
power spectra QML estimates at $\ell < 30$ and 
presented in \cite{Gruppuso2010,Gruppuso2011}.

\begin{table}
%\caption{Cosmological parameters from WMAP 7 year including cosmological birefringence} % title of Table
\centering % used for centering table
\begin{tabular}{c c c} % centered columns (3 columns)
\hline\hline %inserts double horizontal lines
Parameter & WMAP 7 & BoPix plus \\
& likelihood &  WMAP 7 high $\ell$ likelihood\\
%Parameter & Full WMAP 7 likelihood & BoPix + WMAP 7 high $\ell$ likelihood\\
%& full likelihood & & & & & in $T,P$ \\
[0.5ex] % inserts table
%heading
\hline % inserts single horizontal line
100 $\Omega_b h^2$ & $2.226^{+0.057}_{-0.055}$ & $2.217 \pm 0.056$ \\ % inserting body of the table
$\Omega_c h^2$ & $0.1109^{+0.0054}_{-0.0055}$ & $0.1145^{+0.0056}_{-0.0055}$ \\
%$\theta $ & $1.0383 \pm 0.0027$ & $1.0381 \pm 0.0026$ \\ %[1ex] % [1ex] adds vertical space
$\tau$ & $0.0873^{+0.0147}_{-0.0144}$ & $0.0899^{+0.0147}_{-0.0156}$ \\
$n_s$ & $0.964 \pm 0.014$ & $0.957 \pm 0.014$ \\
${\rm log} [10^{10} A_s]$ & $3.191 \pm 0.032$
& $3.138 \pm 0.033$ \\
$\alpha ({\rm rad})$ & $-0.058 < \alpha < 0.025$ & $-0.063 < \alpha < 0.018$ \\ \hline %inserts single line
$\Omega_M$ & $0.270 \pm 0.028$  & $0.290^{+0.030}_{-0.031}$ \\
$H_0$ & $70.4 \pm 2.4$ & $68.9^{+2.4}_{-2.5}$ \\
$\sigma_8$ & $0.805 \pm 0.029$ & $0.823 \pm 0.029$ 
\end{tabular}
\label{marginalized_biri} % is used to refer this table in the text
\caption{Mean parameter values and bounds of the central 68\%-credible intervals
for the cosmological parameters allowing for an effective treatment of cosmological birefringence                          
estimated by the WMAP 7 year full likelihood
(left column) and by the BoPix plus WMAP 7 year high $\ell$ likelihood (right column).
For the angle $\alpha$ defined in Eq. (\ref{C_ll_EB_constant}) the 95\%-credible upper bound is quoted.}
\end{table}

\begin{figure}
\includegraphics[width=9cm,height=5cm,angle=0]{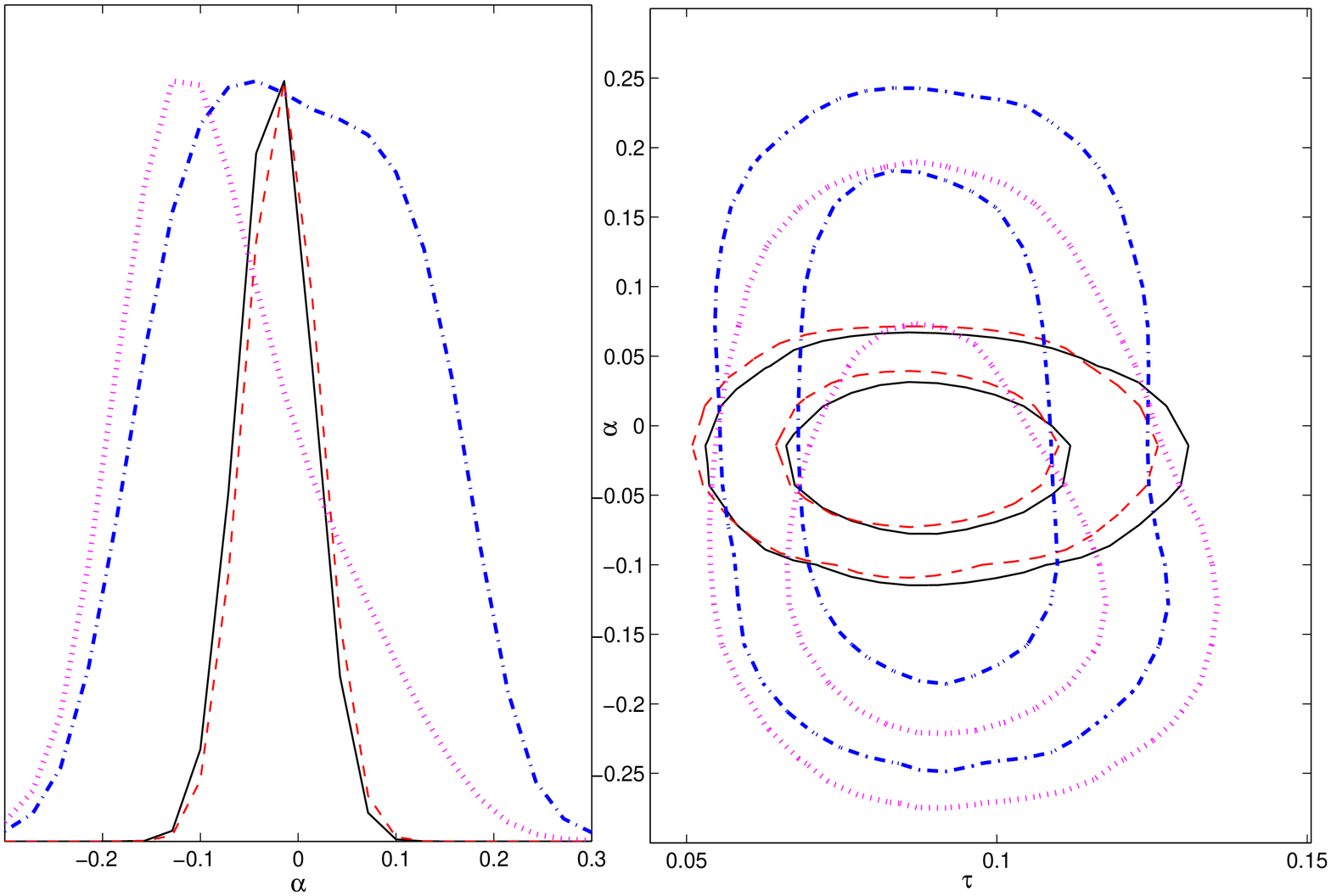}
\label{figuracinque}
\caption{Marginalized posterior probability for $\alpha$ (left panel) and marginalized 
68\%\ and 95\%-credible contours for 
$(\tau \,, \alpha)$ (right panel)
as estimated by the WMAP 7 year full likelihood (dashed red lines)
and by the BoPix plus WMAP 7 year high $\ell$ likelihood (solid black lines). The additional dot-dashed blue line and short-dashed pink 
lines are for the constraints on $\alpha$ from large angular scales only 
obtained by the WMAP 7 year pixel likelihood code and BoPix, respectively.}
\end{figure}

\section{Conclusions}
\label{conclusions}

We have performed an alternative estimate of the cosmological parameters from WMAP 7 year public data, by substituting 
the WMAP 7 low-$\ell$ likelihood with a pixel likelihood code which treats $(T,Q,U)$ at the same resolution without 
any approximation. We have used this code at the HEALPIX resolution $N_{\rm side}=16$ on foreground cleaned public 
data, therefore increasing the 
resolution of the pixel based polarization products used in our extraction of the cosmological parameters with respect 
to the WMAP standard one. 
%Apart from increasing the transition multipole from $\ell=24$ to $\ell=31$ 
%for the high-$\ell$ WMAP 7 year temperature-polarization cross-correlation likelihood, 
We have consistently increased the transition multipole from $\ell=24$ to $\ell=31$
for the high-$\ell$ WMAP 7 year temperature-polarization cross-correlation likelihood and  
included the marginalization over the nuisance parameter $A_{\rm SZ}$. 
%which accounts for the unknown amplitude of the thermal SZ contribution to 
%the small-scale CMB data points assuming the model by \cite{KoSe}. 

With this setting we have found estimates for the cosmological parameters consistent with those obtained by the 
full WMAP 7 year likelihood package, although for some parameters the differences are of half $\sigma$ or more.
These differences between the two low-$\ell$ likelihood treatments 
we find are larger than the WMAP 7 yr likelihood uncertainties from tests on simulations reported 
in \cite{Larson:2010gs}; however, we need to 
keep in mind that our differences between two likelihood treatments are reported for {\em real} data, 
with WMAP 7 year beam/points source corrections and various marginalizations taken fully into account, 
differently from the simulation analysis performed in \cite{Larson:2010gs}. 
%It is also important to stress that the WMAP assumption to have vanishing noise in temperature is explicitly 
%violated by the ILC WMAP 7 yr temperature 
%map digested by the low-$\ell$ likelihood code (the likelihood test on simulations reported in \cite{Larson:2010gs} 
%might also miss the additon of the 1$\mu$K noise per pixel to the low-resolution temperature maps). 
%Another important aspect is that the low resolution polarization 
%data digested by the two likelihood are different since are at different resolution.
The difference between the two best-fit $C_\ell^{TT}$ for $\Lambda$CDM found by the two alternative 
likelihood treatments show a maximum of $4\%$ around at 
$\ell \sim 10$ and oscillate with an amplitude below $1\%$ for $\ell > 100$ 
\footnote{We have checked that either the difference between the two best-fit $C_\ell$ 
or between the estimates of the cosmological parameters decrease when the nuisance parameter 
$A_{\rm SZ}$ is set to zero in both alternative likelihood 
treatments. The net effect of the variation of this foreground parameter, which is unconstrained by the data, 
is to increase the differences between the estimates of the cosmological parameters from the two likelihood 
treatments for the 
$\Lambda$CDM model.}. A $5\%$ 
percent difference is found in the two best-fits for the lensing power spectrum, whereas smaller  
differences are found for temperature-polarization cross-correlation and polarization power spectra.
We have shown how part of the discrepancy, but not all, can 
be ascribed to the monopole/dipole marginalization used in the WMAP 
temperature likelihood and described in \cite{Slosar04}.

On restricting to the $\Lambda$CDM model the most important difference is for the scalar spectral index $n_S$, 
which decrease to 0.956 from the value 0.968 we obtain with the full WMAP 7 yr likelihood code, i.e. a decrease of 
0.86 $\sigma$. This different value for $n_S$ would increase 
the evidence against the Harrison-Zeldovich spectrum from WMAP 7 yr data.
This difference for $n_S$ is consistent with the one between the two best-fit $C_\ell$ and depend only partially 
from the threshold multipole from which the high-$\ell$ TE likelihood starts. 
Other previous alternative likelihood treatments also reported the most important   
discrepancy for the scalar spectral index \citep{Eriksenetal2007,Rudiordetal2007a}.  
A smaller value for $n_S$ with respect to the estimate by the full WMAP 7 year likelihood code, always within 1 $\sigma$, 
is then seen in all the extension of $\Lambda$CDM considered here. 
No major changes are found for the 95 \% credible intervals for the tensor to scalar ratio and 
for the running of the scalar spectral index. A slight degradation 
has been found for the 95 \% credible interval on the neutrino mass. The case of cosmological 
birefringence has been taken as a sensitive test for the two alternative likelihoods, 
whose most relevant difference is the treatment of polarization on large scales. 
%As expected, 
A slight difference on the posterior of the polarization angle 
$\alpha$ has been found when only low resolution data are used, whereas the results 
are fully consistent when the high-$\ell$ TB data are added to both likelihoods.

%We have found a $\Lambda$CDM model 

\section*{Acknowledgements}

We thank Paolo Natoli for comments on the manuscript and for help in the generation of the data set used 
in \cite{Gruppuso2011}, also used here, and Eiichiro Komatsu for useful comments.
We thank Loris Colombo for comparison of our code BoPix with his independent pixel base code BFlike \citep{rocha2010}. 
We thank the Planck CTP and C2 working groups for stimulating and fruitful interactions.
We wish to thank Matteo Galaverni for useful discussion on cosmological birefringence, 
Luca Pagano for useful comments on the WMAP likelihood code and Jan Hamann for useful comments on the manuscript.
We acknowledge the use of the SP6 at CINECA under the agreement LFI/CINECA and of the IASF Bologna cluster.
We acknowledge use of the HEALPix \citep{gorski} software and analysis package for
deriving the results in this paper.  
We acknowledge the use of the Legacy Archive for Microwave Background Data Analysis (LAMBDA). 
Support for LAMBDA is provided by the NASA Office of Space Science. 
Work supported by ASI through ASI/INAF Agreement I/072/09/0 for
the Planck LFI Activity of Phase E2 and by MIUR through PRIN 2009 (grant n. 2009XZ54H2).
%The ASI contract Planck LFI activity of Phase E2 is acknowledged.

%\appendix

\end{document}